\RequirePackage{lineno}
\documentclass[aps,twocolumn,showpacs,byrevtex,prl,reprint]{revtex4-1}

\usepackage{graphicx}
\usepackage{dcolumn}
\usepackage{bm}
\usepackage{rotating}
\usepackage{epstopdf}
\usepackage{color}
\usepackage{verbatim} 
\usepackage{multirow}
\usepackage[abs]{overpic}
\usepackage{amsmath}
\usepackage{amssymb}
\usepackage{xspace}

\newcommand{\PreserveBackslash}[1]{\let\temp=\\#1\let\\=\temp}
\newcolumntype{C}[1]{>{\PreserveBackslash\centering}p{#1}}
\newcolumntype{R}[1]{>{\PreserveBackslash\raggedleft}p{#1}}
\newcolumntype{L}[1]{>{\PreserveBackslash\raggedright}p{#1}}

\newcommand{\LL}{\ell^+\ell^-}
\newcommand{\EE}{e^+e^-}

\newcommand{\too}{\rightarrow}

\def\pb {\ensuremath{\mbox{\,pb}}\xspace}
\def\invpb {\ensuremath{\mbox{\,pb}^{-1}}\xspace}
\def\gev   {\ensuremath{\mbox{\,GeV}\xspace}}

\begin{document}
\graphicspath{{figure/}}
\DeclareGraphicsExtensions{.eps,.png,.ps}
\title{\boldmath Observation of $\EE \too \omega \chi_{c1,2}$ near $\sqrt{s}$ = 4.42 and 4.6 GeV}
\author{
  \begin{small}
    \begin{center}
      M.~Ablikim$^{1}$, M.~N.~Achasov$^{9,e}$, X.~C.~Ai$^{1}$,
      O.~Albayrak$^{5}$, M.~Albrecht$^{4}$, D.~J.~Ambrose$^{44}$,
      A.~Amoroso$^{49A,49C}$, F.~F.~An$^{1}$, Q.~An$^{46,a}$,
      J.~Z.~Bai$^{1}$
% ...
      \\
      \vspace{0.2cm}
      (BESIII Collaboration)\\
      \vspace{0.2cm} {\it
        $^{1}$ Institute of High Energy Physics, Beijing 100049, People's Republic of China\\
% ...
$^{4}$ Bochum Ruhr-University, D-44780 Bochum, Germany\\
        $^{5}$ Carnegie Mellon University, Pittsburgh, Pennsylvania 15213, USA\\
% ...
$^{9}$ G.I. Budker Institute of Nuclear Physics SB RAS (BINP), Novosibirsk 630090, Russia\\
% ...
$^{44}$ University of Rochester, Rochester, New York 14627, USA\\
% ...
$^{46}$ University of Science and Technology of China, Hefei 230026, People's Republic of China\\
% ...
$^{49}$ (A)University of Turin, I-10125, Turin, Italy; (B)University of Eastern Piedmont, I-15121, Alessandria, Italy; (C)INFN, I-10125, Turin, Italy\\
% ...
$^{a}$ Also at State Key Laboratory of Particle Detection and Electronics, Beijing 100049, Hefei 230026, People's Republic of China\\
% ...
$^{e}$ Also at the Novosibirsk State University, Novosibirsk, 630090, Russia\\
% ...
      }\end{center}
    \vspace{0.4cm}
\end{small}
}
\affiliation{}


\begin{abstract}
Based on data samples collected with the BESIII detector operating at
the BEPCII storage ring at center-of-mass energies $\sqrt{s} >$  4.4
GeV, the processes $\EE \too \omega \chi_{c1,2}$ are observed for the
first time. With an integrated luminosity of $1074 \invpb$ near
$\sqrt{s} =$ 4.42 GeV, a significant $\omega \chi_{c2}$ signal is
found, and the cross section is measured to be $(20.9 \pm 3.2 \pm 2.5)\pb$.
With $567 \invpb$ near $\sqrt{s} =$ 4.6 GeV, a clear $\omega
\chi_{c1}$ signal is seen, and the cross section is measured to be
$(9.5 \pm 2.1 \pm 1.3) \pb$, while evidence is found for an $\omega
\chi_{c2}$ signal. The first errors are statistical and the second are
systematic. Due to low luminosity or low cross section at other energies, no significant signals are observed.
In the $\omega \chi_{c2}$ cross section, an enhancement is
seen around $\sqrt{s} =$ 4.42 GeV. Fitting the cross section with a
coherent sum of the $\psi(4415)$ Breit-Wigner function and a phase
space term, the branching fraction
$\mathcal{B}(\psi(4415)\to\omega\chi_{c2})$ is obtained to be of the
order of $10^{-3}$.
\end{abstract}


\maketitle
In recent years, charmonium physics gained renewed strong interest
from both the theoretical and the experimental side, due to the
observation of charmonium-like states, such as
$X(3872)$~\cite{X3872,CDF}, $Y(4260)$~\cite{Y4260,cleo,belle},
$Y(4360)$~\cite{Y4360, Y4660} and $Y(4660)$~\cite{Y4660}. These states
do not fit in the conventional charmonium spectroscopy, and could be exotic
states that lie outside the quark model~\cite{exotic}. Moreover, charged
charmonium-like states $Z_{c}(3900)$~\cite{X5, X6, X7, pi0pi0jpsi},
$Z_{c}(3885)$~\cite{X8, pi0DDstar}, $Z_{c}(4020)$~\cite{X9, X10} and
$Z_{c}(4025)$~\cite{X11, pi0DstarDstar} or their neutral partners were
observed, which might indicate the presence of
new dynamics in this energy region. Searches for new decay modes and
measurements of their line shapes may help us gain a better understanding of
the nature of charmonium(-like) states.

Most recently, BESIII has observed the process $\EE \too \omega
\chi_{c0}$ around $\sqrt{s}$=4.23 GeV~\cite{chunhua}, which has first been
proposed in Ref.~\cite{tang}. As the line shape is incompatible with
that of $Y(4260)$ in $\EE \too \pi^{+}\pi^{-}J/\psi$, the authors of
Ref.~\cite{4s} suggest the excess of $\omega \chi_{c0}$ events due to a
missing charmonium state, while Ref.~\cite{4160} attributes it to the
tail of the $\psi(4160)$. A similar pattern could be expected for the other spin
triplet $P$-wave states $\chi_{c1,2}$.  It is therefore very interesting to search for
$e^+e^- \too \omega\chi_{c0,1,2}$ in the BESIII data collected at
$\sqrt{s} > 4.4$~GeV. The $\omega$-transition may help us to
establish connections between these charmonium(-like) states.

In this Letter, we report on a study of
$\EE\too\omega\chi_{cJ}(J=0,1,2)$ based on the $\EE$ annihilation data
collected with the BESIII detector~\cite{besiii} at five energy points
in the range 4.416 $\leqslant\sqrt{s}\leqslant$ 4.599 GeV. The
integrated luminosity of this data is measured by using Bhabha scattering
with an accuracy of 1.0\%~\cite{luminosity}, and the center-of-mass
energies are measured by using the di-muon process~\cite{ecms}. The
$\chi_{c1,2}$ states are detected via $\chi_{c1,2} \too \gamma
J/\psi$, $J/\psi \too \LL (\ell = e,\mu)$, and the $\omega$ is
reconstructed via the $\omega \too \pi^{+}\pi^{-}\pi^{0}$ decay
mode. For $\EE \too \omega \chi_{c0}$, $\chi_{c0}$ is reconstructed
via its decays to $\pi^{+}\pi^{-}$ or $K^{+}K^{-}$.

\begin{figure}[htbp]
\begin{center}
\begin{overpic}[width=0.23\textwidth]{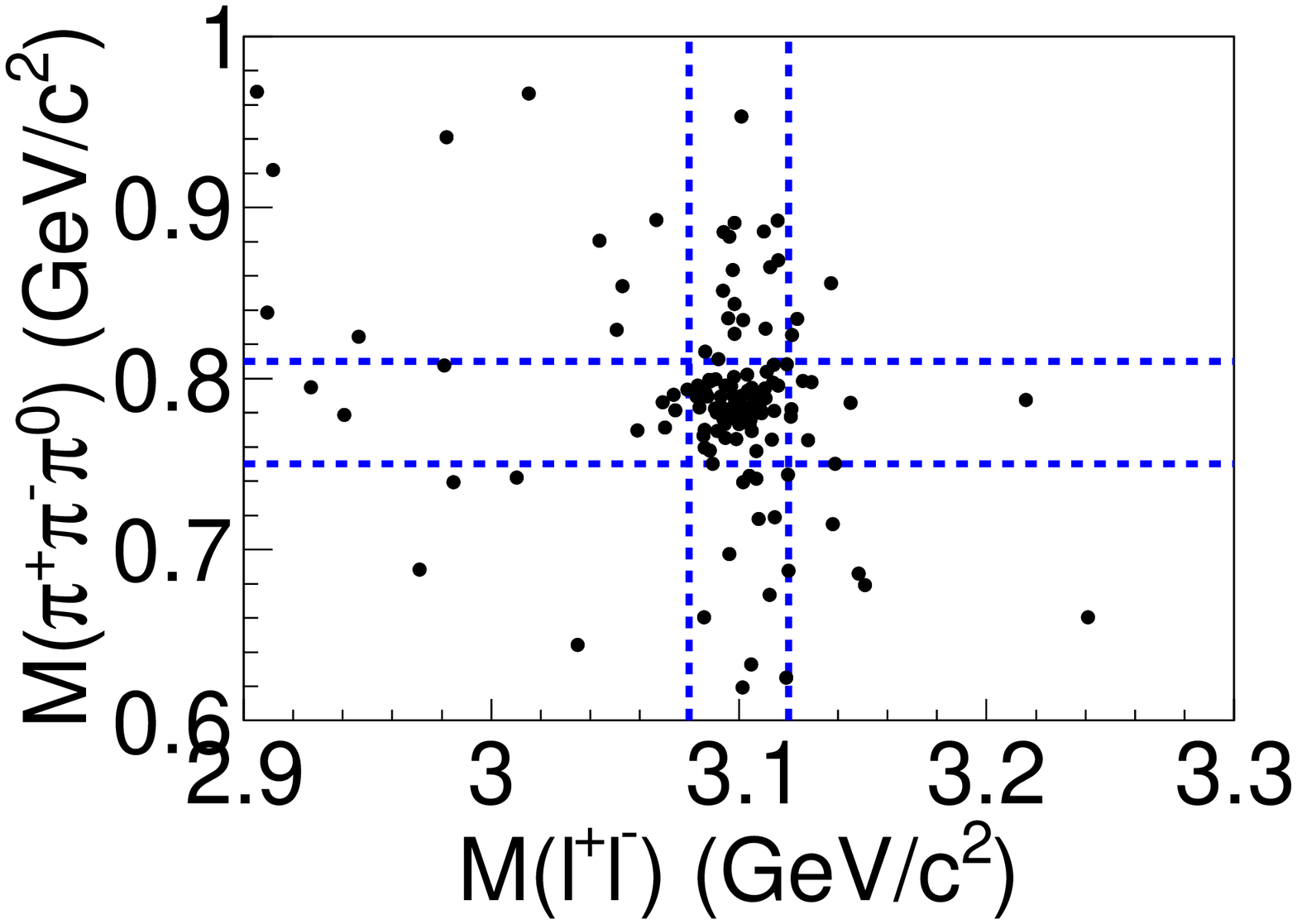}
\put(98,70){(a)}
\end{overpic}
\begin{overpic}[width=0.23\textwidth]{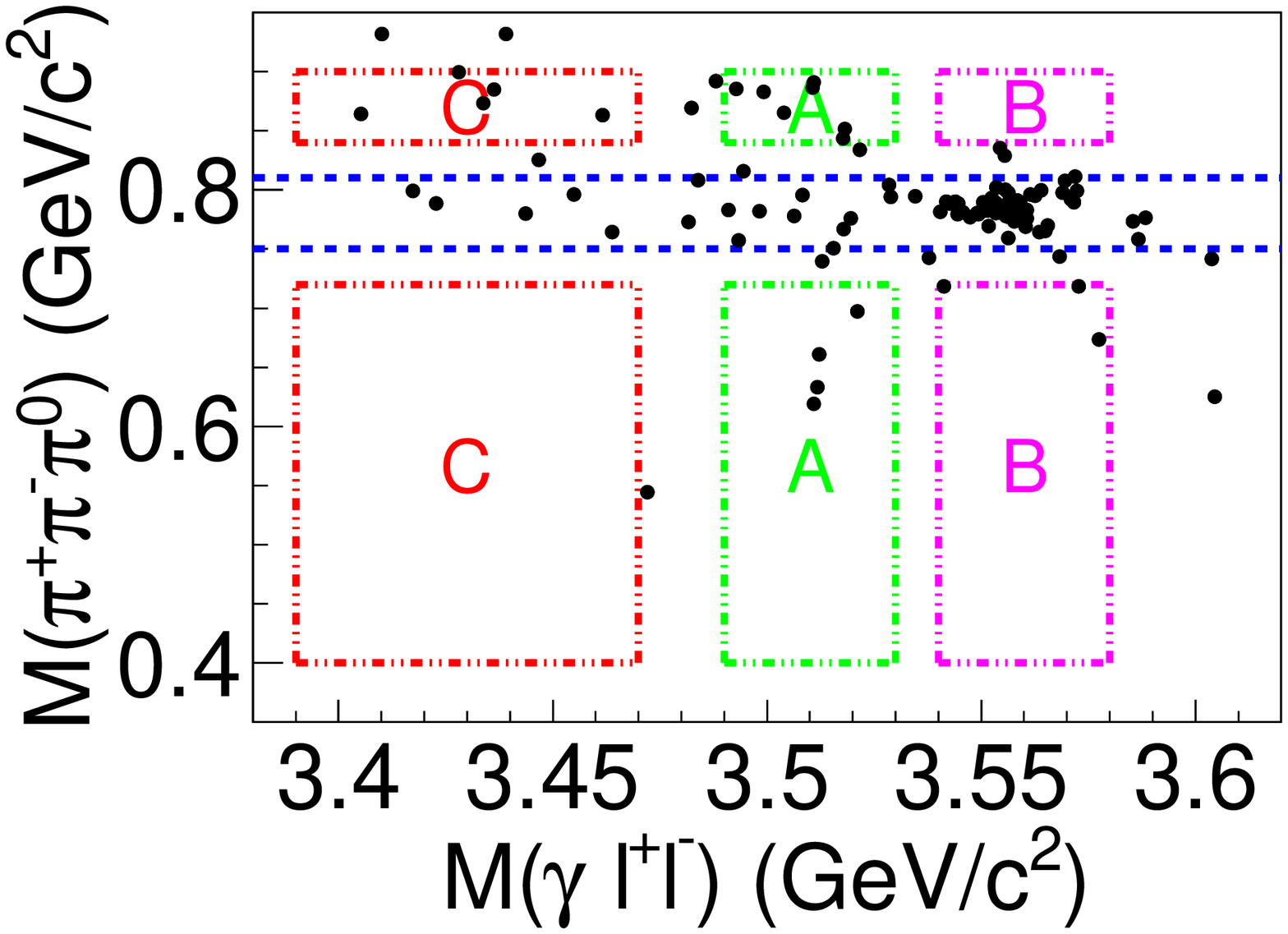}
\put(98,70){(b)}
\end{overpic}
\begin{overpic}[width=0.23\textwidth]{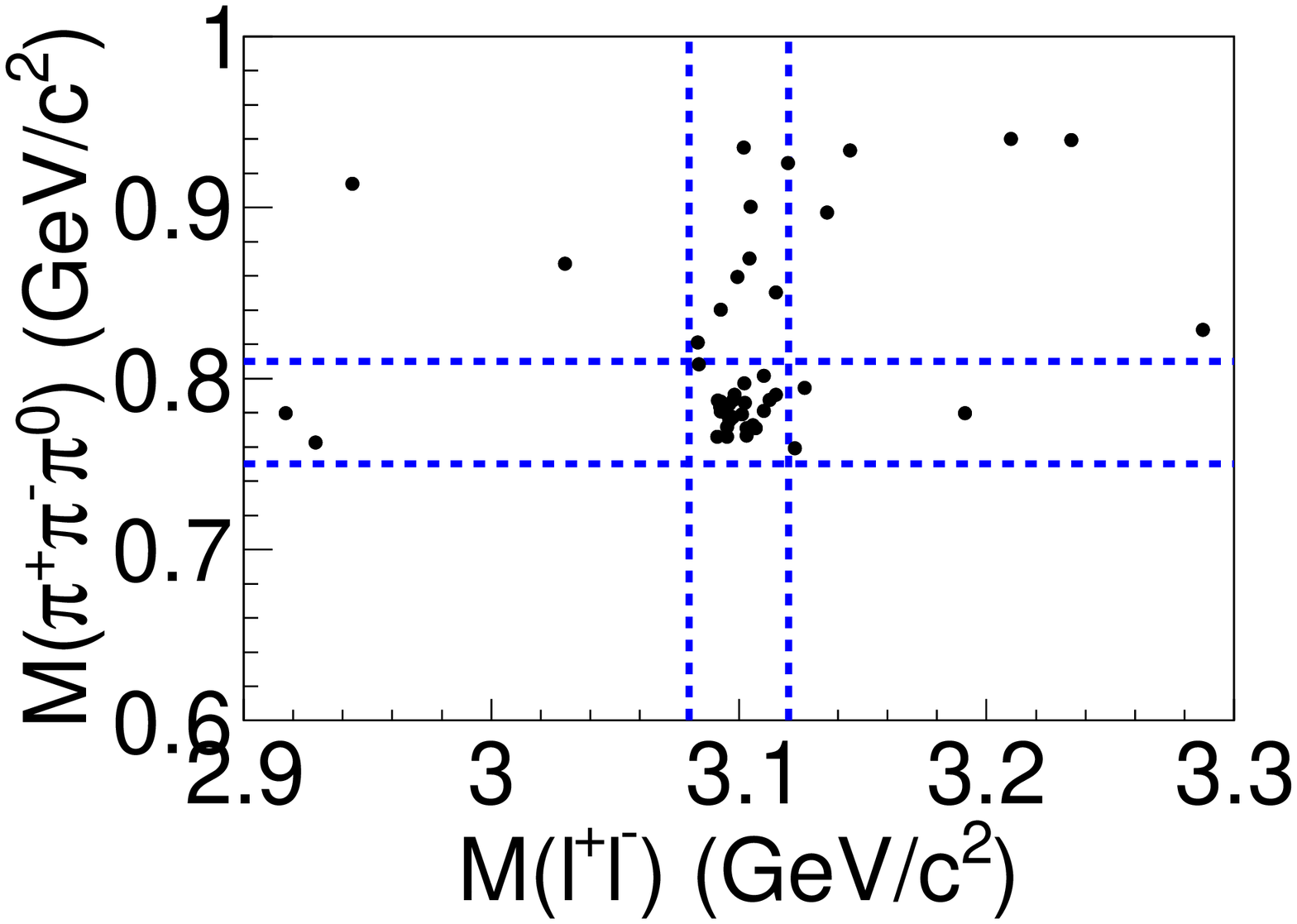}
\put(98,70){(c)}
\end{overpic}
\begin{overpic}[width=0.23\textwidth]{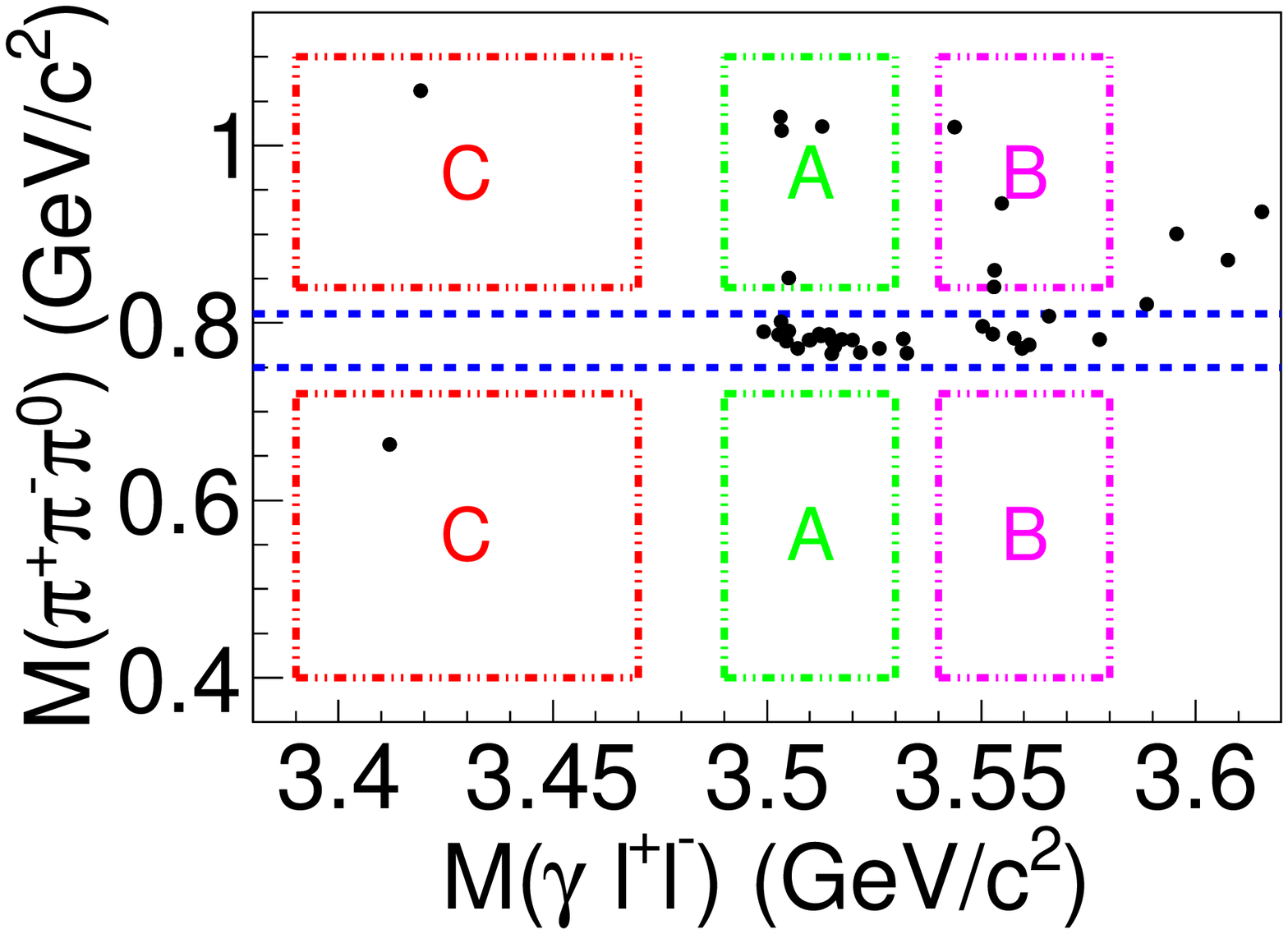}
\put(98,70){(d)}
\end{overpic}
\caption{Scatter plots for data at $\sqrt{s}$ = 4.416 GeV ((a) and
  (b)) and 4.599 GeV ((c) and (d)). Plots (a) and (c) are
  $M(\ell^{+}\ell^{-})$ versus $M(\pi^{+}\pi^{-}\pi^{0})$, the blue
  dashed lines mark the signal region of $\omega$ or $J/\psi$. Plots (b) and
  (d) are $M(\pi^{+}\pi^{-}\pi^{0})$ versus $M(\gamma
  \ell^{+}\ell^{-})$, the blue dashed lines mark the signal region of
  $\omega$, the non-$\omega$ regions (box A,B,C) are used to estimate
  the $\pi^{+}\pi^{-}\pi^{0}\chi_{c1,2}$ events in the $\chi_{c1,2}$
  signal regions.}
\label{fig:scatter}
\end{center}
\end{figure}

Since the final state of the process $\EE \too \omega\chi_{c1,2}$ is
$\gamma\pi^{+}\pi^{-}\pi^{0}\ell^{+}\ell^{-}$, signal candidates must
have exactly four tracks with zero net charge, a $\pi^{0}$ candidate
and a photon. The event selection criteria are the same as described in
Ref.~\cite{chunhua}. A five constraint (5C)-kinematic fit is performed
constraining the total four-momentum of the final state to
the initial four-momentum of the colliding beams, and the
invariant mass of the two photons from $\pi^{0}$ is constrained to
the nominal $\pi^{0}$ mass. The $\chi^{2}_{5C}$ of candidate events is
required to be less than 60. The scatter plots of
$M(\ell^{+}\ell^{-})$ versus $M(\pi^{+}\pi^{-}\pi^{0})$ after the
above requirements are
shown in Fig.~\ref{fig:scatter} ((a) and (c)) for data at $\sqrt{s}$ =
4.416 GeV and 4.599 GeV. Clear signals are seen in the $\omega$ and
$J/\psi$ signal regions, which are defined as $0.75 \leqslant
M(\pi^{+}\pi^{-}\pi^{0}) \leqslant 0.81\,\text{GeV}/c^{2}$
for $\omega$ and $3.08\leqslant M(\ell^{+}\ell^{-}) \leqslant 3.12 \,\text{GeV}/c^{2}$ for $J/\psi$,
respectively. The mass resolution for $J/\psi$ is found to be 8 MeV/$c^{2}$
in Monte Carlo (MC) simulations. The scatter plots of
$M(\pi^{+}\pi^{-}\pi^{0})$ versus $M(\gamma \ell^{+}\ell^{-})$ after
the $J/\psi$ requirement are shown in Fig.~\ref{fig:scatter} ((b) and
(d)). The signal regions of $\chi_{c1}$ and $\chi_{c2}$ are set to be
[3.49, 3.53] and [3.54, 3.58] GeV/$c^{2}$, respectively, and the
region [3.39, 3.47] GeV/$c^{2}$ is taken as the $\chi_{c1,2}$
sideband. Clear accumulations of events can be seen in the $\chi_{c2}$
signal region at $\sqrt{s}$ = 4.416 GeV and in the $\chi_{c1}$ signal
region at $\sqrt{s}$ = 4.599 GeV.

The main backgrounds are found to be $\EE \too
\pi^{+}\pi^{-}\pi^{0}\chi_{c1,2}$, where $\pi^{+}\pi^{-}\pi^{0}$ are
of non-resonant origin. The $\pi^{+}\pi^{-}\pi^{0}\chi_{c1,2}$
background will produce a peak in the $\chi_{c1,2}$ signal region. The
non-$\omega$ regions (box A, B, C), as shown in
Fig.~\ref{fig:scatter}, are used to estimate the background. The
number of $\pi^{+}\pi^{-}\pi^{0}\chi_{c1,2}$ events in the
$\chi_{c1,2}$ signal regions can be calculated by
$n^{bkg}_{1,2}=f\cdot(n_{A,B}-0.5n_{C})$, where $n_{A}, n_{B}, n_{C}$
are the numbers of events in boxes $A$, $B$, and $C$, and $f$ is a normalization
factor. To estimate the normalization factor $f$, we use the
phase-space (PHSP) generator to simulate
$\pi^{+}\pi^{-}\pi^{0}\chi_{c1,2}$ events at $\sqrt{s}$ = 4.416 and
4.599 GeV.

Other possible backgrounds come from $\EE \too \eta' J/\psi$ with
$\eta' \too \gamma \omega$, $\EE \too \pi^{+}\pi^{-}\psi'$ with
$\psi' \too \pi^{0}\pi^{0}J/\psi$ or $\gamma\chi_{c1,2}$, and
$\EE \too \pi^{0}\pi^{0}\psi'$ with $\psi' \too \pi^{+}\pi^{-}J/\psi$. All
these backgrounds will not produce peaks in the signal regions, and
their contribution is estimated to be negligible.

\begin{figure}[htbp]
\begin{center}
\includegraphics[width=0.23\textwidth]{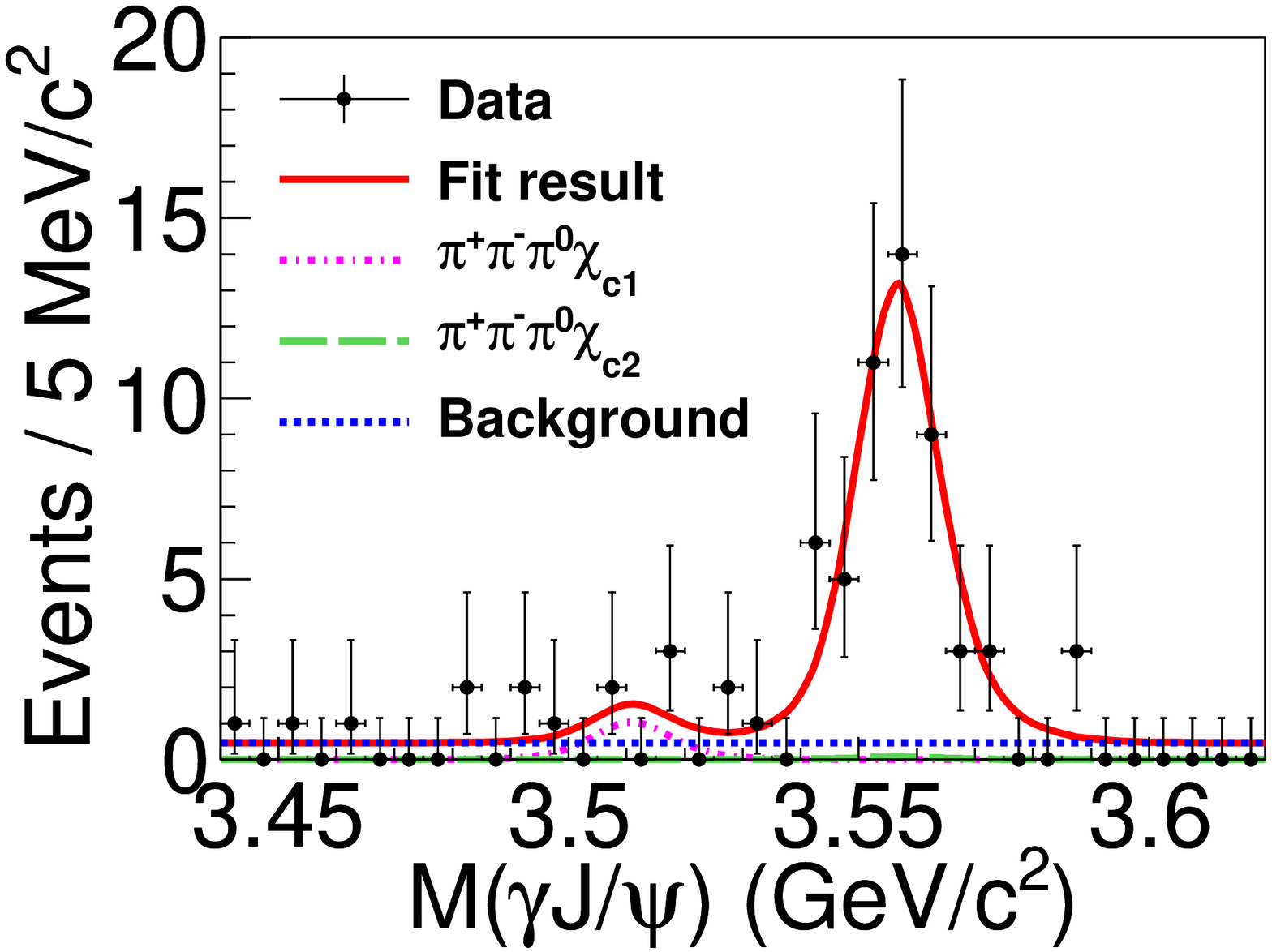}
\includegraphics[width=0.23\textwidth]{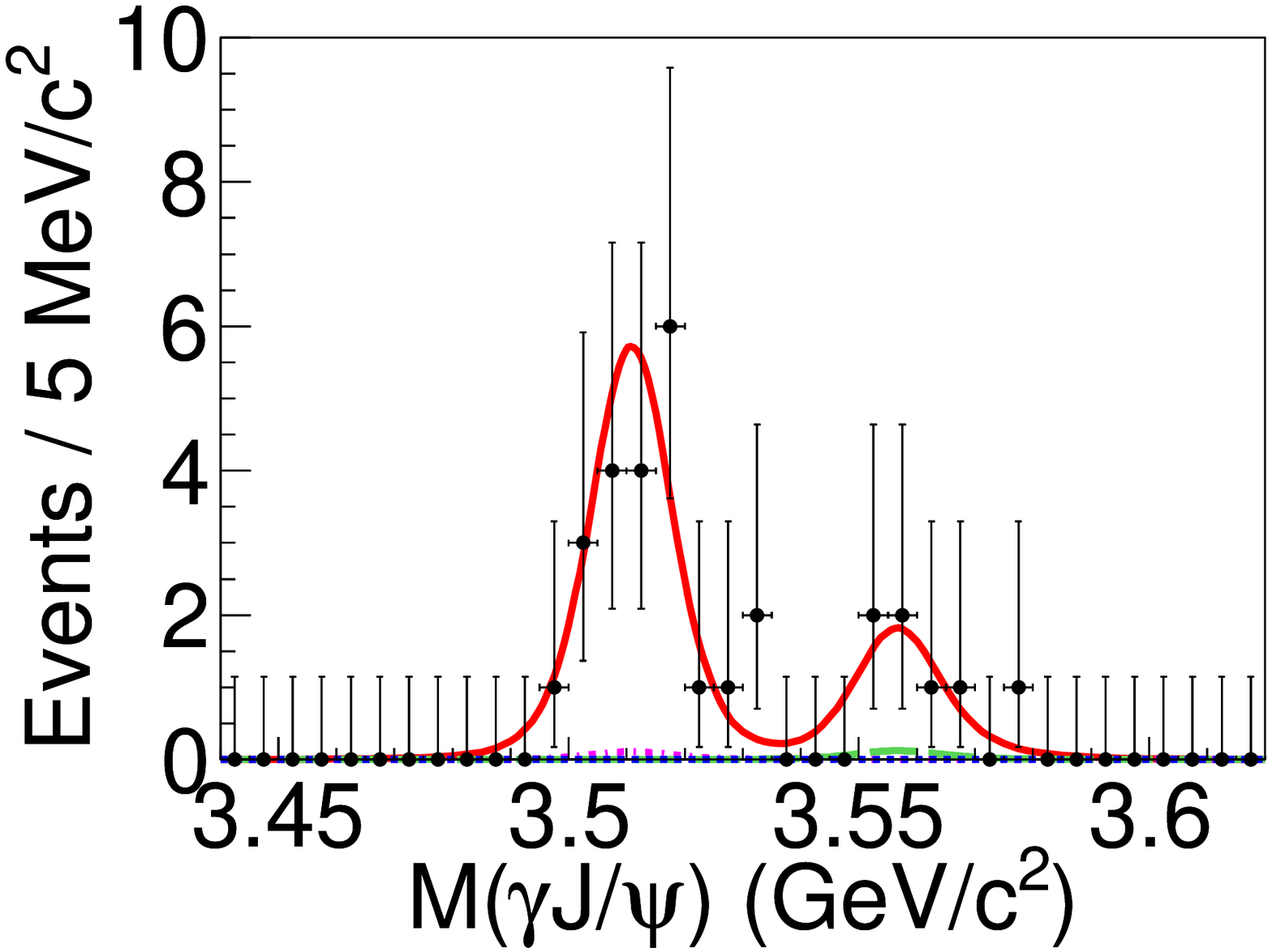}
\caption{Fits to the invariant mass $M(\gamma J/\psi)$ distributions
  for data at $\sqrt{s} = 4.416\gev$ (left) and $4.599\gev$ (right). The
  red solid curves are the fit results. The magenta dashed-dotted
  curves and green long-dashed curves show the
  $\pi^{+}\pi^{-}\pi^{0}\chi_{c1}$ and
  $\pi^{+}\pi^{-}\pi^{0}\chi_{c2}$ peaking backgrounds, the blue dashed
  curves represent the flat background.}
\label{fig:fitresult}
\end{center}
\end{figure}

Figure~\ref{fig:fitresult} shows the $M(\gamma J/\psi)$ distributions
at $\sqrt{s} = 4.416$ and 4.599 GeV for events in the $J/\psi$ and
$\omega$ signal region. Significant $\chi_{c2}$ signals at
$\sqrt{s}=4.416 \gev$ and $\chi_{c1}$ signals at $\sqrt{s}=4.599 \gev$
are visible. Unbinned maximum likelihood fits are performed
to measure the signal yields. The signal shapes are
determined from signal MC samples. The shapes of the peaking
background are determined by the $\pi^{+}\pi^{-}\pi^{0}\chi_{c1,2}$ MC
sample, and the magnitudes are fixed at the expectation based on the
non-$\omega$ region as mentioned above. The non-peaking backgrounds are
described with a constant. The fit results are shown in
Fig.~\ref{fig:fitresult}. For data at $\sqrt{s} = 4.416\gev$, the
$\omega\chi_{c1}$ signal yield is $0.0^{+3.7}_{-0.0}$, and the
$\omega\chi_{c2}$ signal yield is $49.3\pm7.5$. The statistical
significance of the $\chi_{c2}$ signal is 10.4 $\sigma$ by comparing
the difference of log-likelihood values
($\Delta(\rm{ln}\mathcal{L})=54.0$) with or without the $\chi_{c2}$ signal
in the fit and taking into account the change of the number of
degrees-of-freedom ($\Delta \text{ndf}=1$). For data at $\sqrt{s} = 4.599
\gev$, the $\omega\chi_{c1}$ signal yield is $21.1\pm4.7$ with a
statistical significance of 7.4 $\sigma$
($\Delta(\rm{ln}\mathcal{L})=27.5$, $\Delta \text{ndf}=1$), and the
$\omega\chi_{c2}$ signal yield is $7.0^{+3.2}_{-2.5}$ with a
statistical significance of 3.8 $\sigma$
($\Delta(\rm{ln}\mathcal{L})=7.1$, $\Delta \text{ndf}=1$). The detailed
information can be found in Table~\ref{tab:crosssection}.

\begin{table*}[htbp]
\begin{center}
  \caption{Results on $e^+e^-\to \omega \chi_{cJ}(J=0,1,2)$. Shown
    in the table are the channels, the center-of-mass energy, the
    integrated luminosity $\mathcal{L}$, product of radiative
    correction factor, vacuum polarization factor, branching fraction
    and efficiency,
    $\mathcal{D}=(1 + \delta) \frac{1}{|1-\Pi|^{2}}
    (\epsilon_{e}\mathcal{B}_{e} + \epsilon_{\mu}\mathcal{B}_{\mu})
    \mathcal{B}_{1}$
    for $\omega\chi_{c1,2}$ and
    $\mathcal{D}=(1+\delta)\frac{1}{|1-\Pi|^{2}}(\epsilon_{\pi}\mathcal{B}(\chi_{c0}\to\pi^+\pi^-)+\epsilon_{K}\mathcal{B}(\chi_{c0}\to
    K^+K^-))\mathcal{B}(\omega\to\pi^{+}\pi^{-}\pi^{0})\mathcal{B}(\pi^{0}\to\gamma\gamma)$
    for $\omega\chi_{c0}$, number of observed events $N^{\rm {obs}}$,
    number of estimated background events $N^{\rm bkg}$, number of
    signal events $N^{\rm sig}$ determined as described in the text, Born cross section
    $\sigma^{\rm B}$(or upper limit at 90$\%$ C.L.) at each energy
    point. Here the first errors are statistical, and the second
    systematic. $N^{\rm sig}$ for $\omega\chi_{c1,2}$ at $\sqrt{s}$ =
    4.416 and 4.599 GeV is taken from the fit. Dash means that the
    result is not applicable.}
\label{tab:crosssection}
\begin{tabular}{cccccccc}
  \hline
  \hline
  Channel & $\sqrt{s}$ (GeV) & $\mathcal{L}$(pb$^{-1}$) & $\mathcal{D}~(\times10^{-3})$  &  $N^{\rm {obs}}$ &  $N^{\rm bkg}$  & $N^{\rm sig}$ &  $\sigma^{\rm B}$ (pb) \\
  \hline
  $\omega\chi_{c0}$ & 4.416 & 1074  & 2.35  &  52   & $54.6\pm9.4$         & $0.0^{+9.5}_{-0.0}$     & $<7$    \\
                    & 4.467 & 110   & 2.14  &  13   & $9.4\pm1.7$          & $3.6^{+4.2}_{-3.6}$     & $<49$   \\
                    & 4.527 & 110   & 2.14  &  7    & $4.1\pm1.3$          & $2.9^{+3.0}_{-2.6}$     & $<37$   \\
                    & 4.574 & 48    & 1.83  &  3    & $1.4\pm0.7$          & $1.6^{+2.0}_{-1.6}$     & $<65$   \\
                    & 4.599 & 567   & 3.00  &  25   & $21.1\pm3.1$         & $3.9^{+5.8}_{-3.9}$     & $<9$    \\
  \hline
  $\omega\chi_{c1}$ & 4.416 & 1074  & 3.89  & 10    & $6.3^{+2.7}_{-1.9}$  & $0.0^{+3.7}_{-0.0}$     & $<3$   \\
                    & 4.467 & 110   & 3.72  &  3    & $0.0^{+0.6}_{-0.0}$  & $3.0^{+3.0}_{-1.6}$     & $<18$  \\
                    & 4.527 & 110   & 3.71  &  0    & $0.0^{+0.6}_{-0.0}$  & $0.0^{+1.3}_{-0.0}$     & $<5$   \\
                    & 4.574 & 48    & 3.77  &  3    & $0.0^{+0.6}_{-0.0}$  & $3.0^{+3.0}_{-1.6}$     & $<39$  \\
                    & 4.599 & 567   & 3.91  &  $-$  & $-$                  & 21.1 $\pm$ 4.7          & 9.5 $\pm$ 2.1 $\pm$ 1.3          \\
  \hline
  $\omega\chi_{c2}$ & 4.416 & 1074  & 2.20  &  $-$  & $-$                  & 49.3 $\pm$ 7.5          & 20.9 $\pm$ 3.2 $\pm$ 2.5          \\
                    & 4.467 & 110   & 2.16  &  4    & $0.0^{+0.6}_{-0.0}$  & $4.0^{+3.2}_{-1.9}$     & $<36$  \\
                    & 4.527 & 110   & 2.18  &  0    & $0.0^{+0.6}_{-0.0}$  & $0.0^{+1.3}_{-0.0}$     & $<9$   \\
                    & 4.574 & 48    & 2.16  &  2    & $0.0^{+0.6}_{-0.0}$  & $2.0^{+2.7}_{-1.3}$     & $<53$  \\
                    & 4.599 & 567   & 2.30  &  7    & $0.5^{+0.9}_{-0.4}$  & $7.0^{+3.2}_{-2.5}$     & $<11$  \\
\hline
\hline
\end{tabular}
\end{center}
\end{table*}

Due to the limited integrated luminosity, the $\omega \chi_{c1,2}$ signals at the
other energy points ($\sqrt{s}= 4.467, 4.527$ and $4.574\gev$) are not
significant, and upper limits at the 90$\%$ C.L. are derived. The signal
yields are obtained by counting events in the $\chi_{c1,2}$ signal
regions and subtracting the backgrounds which are estimated from the
$\chi_{c1,2}$ sidebands. The peaking backgrounds here are
negligible. For the $\omega \chi_{c0}$ decay mode, signals are not
significant at any of the energy points. We construct a likelihood
function by assuming that the observed events follow a Poisson distribution
and the background events follow a Gaussian distribution, where the
signal yields are limited to be positive. From the likelihood
distribution, the signal yields and uncertainties are determined.

The Born cross section is calculated with
\begin{equation}
\small
    \sigma^{B}(\EE \rightarrow \omega \chi_{c1,2}) = \frac{N^\text{sig}}{\mathcal{L} (1 + \delta) \frac{1}{|1-\Pi|^{2}} (\epsilon_{e}\mathcal{B}_{e} + \epsilon_{\mu}\mathcal{B}_{\mu}) \mathcal{B}_{1} } ,
\end{equation}
where $N^\text{sig}$ is the number of signal events, $\mathcal{L}$ is the
integrated luminosity, (1 + $\delta$) is the radiative correction
factor obtained from a Quantum Electrodynamics (QED)
calculation~\cite{QED, KKMC} using the measured cross section as input
and iterated until the results converge; $\frac{1}{|1-\Pi|^{2}}$ is
the vacuum polarization factor which is taken from a QED calculation
with an accuracy of $0.5\%$~\cite{vacuum}; $\epsilon_{e}$
($\epsilon_{\mu}$) is the global selection efficiency for the $\EE \too \omega\chi_{c1,2}, \chi_{c1,2} \too \gamma J/\psi, J/\psi \too \EE$
($\mu^{+}\mu^{-}$) decay mode, $\mathcal{B}_{e}$ ($\mathcal{B}_{\mu}$)
is the branching fraction $\mathcal{B}(J/\psi \too \EE)$
($\mathcal{B}(J/\psi \too \mu^{+}\mu^{-})$), $\mathcal{B}_{1}$ is the
product branching fraction
$\mathcal{B}(\chi_{c1,2} \too \gamma J/\psi) \times \mathcal{B}(\omega
\too \pi^{+}\pi^{-}\pi^{0}) \times \mathcal{B}(\pi^{0} \too
\gamma\gamma)$.
The Born cross section (or its upper limit) at each energy point for
$\EE \too \omega\chi_{cJ}$ is listed in Table~\ref{tab:crosssection}.

For the energy points where the signals are not
significant, the upper limits on the cross sections are
provided. The upper limit is calculated by using a frequentist method
with unbounded profile likelihood, which is implemented by the
package {\sc trolke}~\cite{Rolke} in the {\sc root} framework. The
number of the background events is assumed to follow a Poisson
distribution, and the efficiency is assumed to have Gaussian
uncertainties. In order to consider the systematic uncertainty in the upper limit
calculation, we use the denominator in Eq.(1) as an effective
efficiency as implemented in {\sc trolke}.

The systematic uncertainties on the Born cross section measurement
mainly originate from the detection efficiency, the radiative
corrections, the fit procedure, the branching fractions, and the
luminosity measurement.

The uncertainty in the tracking efficiency is $4.0\%$ for both $\EE$ and
$\mu^{+}\mu^{-}$ decay modes ($1.0\%$ per track)~\cite{chunhua}. The uncertainty in
photon reconstruction is $1.0\%$ per photon, obtained by studying the
$J/\psi \too \rho^{0}\pi^{0}$ decay~\cite{rhopi}.

In order to estimate the uncertainty caused by the angular
distribution, the $\omega$ helicity angular distribution is set to
$1 \pm \cos^{2}\theta_{1}$ (where $\theta_{1}$ is the polar angle of
$\omega$ in the $\EE$ rest frame with the $z$ axis pointing in the
electron beam direction) in the generator instead of the PHSP model,
and the photon (from $\chi_{c1,2}$) helicity angular distribution is
also set to $1 \pm \cos^{2}\theta_{2}$ (where $\theta_{2}$ is the polar
angle of the photon in the $\chi_{c1,2}$ rest frame, with the $z$
axis pointing in the $\omega$ direction) in the generator instead of the
PHSP model. The maximum change in the MC efficiencies is taken as the
systematic uncertainty.

In the analysis, the helix parameters for simulated charged tracks have been
corrected so that the MC simulation matches the momentum spectra of
the data well~\cite{corr}. The correction factors for $\pi$, $e$ and
$\mu$ are obtained by using control samples
$\EE \too \pi^{+}\pi^{-}J/\psi$, $J/\psi \too \EE$ and
$\mu^{+}\mu^{-}$, respectively. The difference in MC efficiency
between results obtained with and without the correction is taken as
the systematic uncertainty.

The line shapes of $\EE \too \omega \chi_{c1,2}$ will affect the radiative
correction factor and the efficiency. The uncertainty is estimated by
varying the line shapes of the cross section in the generator from the
measured cross section to the $Y(4660)$ Breit-Wigner (BW) shape for
$\omega \chi_{c1}$ and to the $\psi(4415)$ BW shape for $\omega
\chi_{c2}$.
The change in the final result between the two line shapes is taken as
the uncertainty from the radiative correction factor.

In the nominal fit, the fit range is taken from 3.44 to 3.62
GeV/$c^{2}$. The uncertainty from the fit range is obtained by
varying the limits of the fit range by $\pm0.025 \gev/c^{2}$. The
systematic uncertainty caused by the flat background shape is estimated by
changing the background shape from a constant to a first-order
polynomial. To estimate the uncertainty caused by the peaking
background, we vary the number of the peaking background events by one
standard deviation in the fit, and cite the larger difference of the
cross sections from the nominal values as the systematic uncertainty.

The luminosity is measured using Bhabha events with an uncertainty of
$1.0\%$~\cite{luminosity}. The branching fractions $\mathcal{B}_{e}$,
$\mathcal{B}_{\mu}$, and $\mathcal{B}_{1}$ are taken from the world
average~\cite{pdg}, and their uncertainties are considered in the systematic
uncertainty. The $J/\psi$ mass window requirement has been studied in
Ref.~\cite{jpsimasswindow}, and a 1.6$\%$ systematic uncertainty is
assigned. The uncertainty due to the cross feed between the $\pi^{+}\pi^{-}$
and $K^{+}K^{-}$ modes is estimated by using the signal MC samples.
\begin{table*}[htbp]
\begin{center}
\caption{Relative systematic uncertainties for the luminosity,
  efficiency, line shape, fit procedure and branching
  fractions (in units of $\%$). The first value in brackets is for
  $\omega\chi_{c0}$, the second for $\omega\chi_{c1}$, and the third
  for $\omega\chi_{c2}$. Dash means that the result is not applicable.
}
\label{tab:sumerror}
\begin{tabular}{c  c  c  c  c  c }
  \hline
  \hline
  Source/$\sqrt{s}$ & 4.416 & 4.467 & 4.527 & 4.574 & 4.599                                \\
  \hline
  Luminosity            & 1.0         & 1.0          & 1.0          & 1.0          & 1.0          \\
  Tracking              & 4.0         & 4.0          & 4.0          & 4.0          & 4.0          \\
  Photon                & (2.0, 3.0, 3.0)         & (2.0, 3.0, 3.0)          & (2.0, 3.0, 3.0)          & (2.0, 3.0, 3.0)          & (2.0, 3.0, 3.0)          \\
  $J/\psi$ mass window  & ($-$, 1.6, 1.6)         & ($-$, 1.6, 1.6)          & ($-$, 1.6, 1.6)          & ($-$, 1.6, 1.6)          & ($-$, 1.6, 1.6)          \\
  Kinematic fit         & (1.3, 2.0, 2.1)  & (1.3, 1.9, 1.6)   & (1.1, 2.0, 1.8)   & (0.3, 2.0, 1.7)   & (0.5, 2.3, 2.0)   \\
  Angular distribution  & (2.5, 6.0, 6.1)  & (3.2, 7.2, 8.3)   & (2.5, 9.6, 8.0)   & (3.5, 9.3, 10.1)  & (1.7, 11.0, 10.3) \\
  Line shape            & (7.1, 1.2, 1.7)  & (13.4, 1.0, 3.6)   & (7.3, 0.5, 1.3)   & (5.6, 0.9, 1.3)   & (10.9, 1.0, 2.9)   \\
  Fit Range             & ($-$, $-$, 3.9)    & $-$            & $-$            & $-$            & ($-$, 0.1, $-$)     \\
  Flat background       & ($-$, $-$, 4.5)    & $-$            & $-$            & $-$            & ($-$, 0.0, $-$)     \\
  Peaking background    & ($-$, $-$, 4.1)    & $-$            & $-$            & $-$            & ($-$, 1.9, $-$)     \\
  Cross feed            & (1.4, $-$, $-$)    & (1.7, $-$, $-$)    & (4.1, $-$, $-$)   & (7.7, $-$, $-$)   & (8.0, $-$, $-$)  \\
  $\mathcal{B}_{e}$, $\mathcal{B}_{\mu}$ & ($-$, $0.6$, $0.6$) & ($-$, $0.6$, $0.6$) & ($-$, $0.6$, $0.6$) & ($-$, $0.6$, $0.6$) & ($-$, $0.6$, $0.6$)  \\
  $\mathcal{B}_{1}$     & (3.8, 3.6, 3.7)  & (3.8, 3.6, 3.7)   & (3.8, 3.6, 3.7)   & (3.8, 3.6, 3.7)   & (3.8, 3.6, 3.7)    \\
  \hline
  Sum                   & (9.8, 9.2, 11.8) & (15.2, 10.0, 11.3) & (10.6, 11.8, 10.6) & (11.4, 11.6, 12.3) & (14.9, 13.2, 12.7)  \\
  \hline
  \hline
\end{tabular}
\end{center}
\end{table*}

Table~\ref{tab:sumerror} summarizes all systematic uncertainties
of the processes $\EE \too \omega\chi_{cJ}$, where the first values in
brackets are for $\omega\chi_{c0}$, the second for $\omega\chi_{c1}$,
and the third for $\omega\chi_{c2}$. The overall systematic
uncertainties are obtained as the quadratic sum of all the sources of
systematic uncertainties, assuming they are independent.

\begin{figure}[htbp]
\begin{center}
\includegraphics[width=0.33\textwidth]{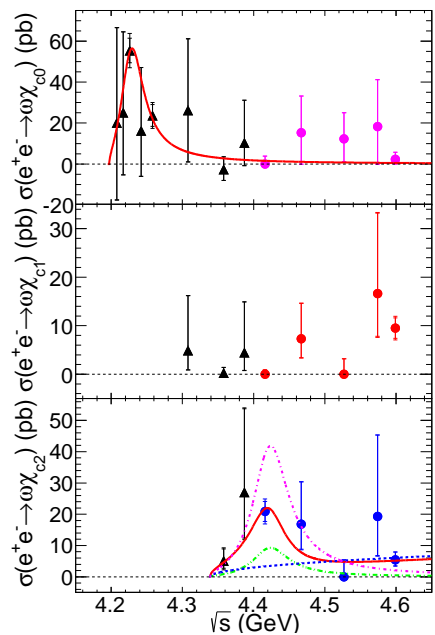}
\caption{ Measured Born cross section (center value) for
  $\EE \too \omega \chi_{cJ}(J=0,1,2)$ as a function of the center of
  mass energy. The top plot is for $\EE \too \omega \chi_{c0}$, the
  middle plot for $\EE \too \omega \chi_{c1}$ and the bottom plot for
  $\EE \too \omega \chi_{c2}$, where the smaller errors are
  statistical only and the larger errors are the quadratic sum of the
  statistical and systematic errors. The triangle black points are
  from Ref.~\cite{chunhua} and others are from this analysis. The
  $\sigma(\EE \too \omega \chi_{c0})$ is fitted with a resonance(solid
  curve) in the top plot. $\sigma(\EE \too \omega \chi_{c2})$ is
  fitted with the coherent sum of the $\psi(4415)$ BW function and a
  phase-space term. The solid curve shows the fit result, the blue
  dashed curve is the phase-space term, which is almost the same for
  the two solutions. The purple dash-dotted curve is the destructive
  solution and the green dash-double-dotted curve is the constructive
  one.}
\label{fig:crosssectionall}
\end{center}
\end{figure}

In Fig.~\ref{fig:crosssectionall}, we compare the line shapes of the
Born cross sections for $\EE \too \omega \chi_{cJ}$, where the Born
cross sections for $\EE \too \omega \chi_{cJ}$ at $\sqrt{s} < 4.4$ GeV
are from Ref.~\cite{chunhua}. Enhancements can be seen in the line
shapes; in the following, we try to fit line shapes. The cross section
of $\EE \too \omega \chi_{c0}$ with the addition of higher energy
points is refitted with a phase-space modified BW
function~\cite{chunhua}, and the fit results for the structure
parameters are
$\Gamma_{ee}\mathcal{B}(\omega\chi_{c0})=(2.8\pm 0.5\pm 0.4)\,\text{eV}$,
$M=(4226\pm 8\pm 6)$~MeV/$c^2$, and $\Gamma_t=(39\pm 12\pm 2)$~MeV,
which are almost unchanged. In the $\EE \too \omega \chi_{c2}$ cross
section, an enhancement is seen around 4.416 GeV, so we use a coherent
sum of the $\psi(4415)$ BW function and a phase space term
\begin{equation}
\footnotesize
    \sigma(\sqrt{s}) = \left|{\frac{\sqrt{12\pi\Gamma_{ee}\mathcal{B}(\omega \chi_{c2})\Gamma_{t}}}{s-M^{2}+iM\Gamma_{t}}\sqrt{\frac{\Phi(\sqrt{s})}{\Phi(M)}}e^{i\phi}+A\sqrt{\Phi(\sqrt{s})}}\right|^{2} ,
\end{equation}
to fit the cross section, where $M$, $\Gamma_t$, $\Gamma_{ee}$ are
mass, total width, $\EE$ partial width for $\psi(4415)$, and are fixed
to the known $\psi(4415)$ parameters~\cite{pdg},
$\mathcal{B}(\omega\chi_{c2})$ is the branching fraction of
$\psi(4415)\to \omega\chi_{c2}$, $\Phi(\sqrt{s}) = p/\sqrt{s}$ is the
phase space factor for an $S$-wave two-body system, where $p$ is the
$\omega$ momentum in the $e^+e^-$ center-of-mass frame, $\phi$ is the
phase angle, and $A$ is the amplitude for the phase-space term. Two
solutions are obtained with the same fit quality, the constructive
solution is $\phi=124^{\circ}\pm35^{\circ}$,
$\mathcal{B}(\omega\chi_{c2})=(1.4\pm0.5)\times10^{-3}$; the
destructive one is $\phi=-105^{\circ}\pm15^{\circ}$,
$\mathcal{B}(\omega\chi_{c2})=(6\pm1)\times10^{-3}$. The goodness of
fit is $\chi^{2}/ndf=4.6/4$.

In summary, using data samples collected at $\sqrt{s} > 4.4$ GeV, the
processes $\EE \too \omega \chi_{c1,2}$ are observed. With an integrated
luminosity of $1074 \invpb$ near $\sqrt{s} = 4.42\gev$, a significant
$\omega \chi_{c2}$ signal is seen, and the cross section is measured
to be $(20.9 \pm 3.2 \pm 2.5) \pb$, where the first uncertainty is
statistical and the second is systematic. Near $\sqrt{s} = 4.6 \gev$
a clear $\omega \chi_{c1}$ signal is observed in $567 \invpb$ of data, with
a cross section of $(9.5 \pm 2.1 \pm 1.3) \pb$; evidence for an
$\omega \chi_{c2}$ signal is found. The $\omega \chi_{c1,2}$ signals
at other energies and the $\omega \chi_{c0}$ signals are not
significant, the upper limits on the Born cross section at 90\%
C.L. are calculated. Interesting line shapes are observed for
$\omega \chi_{cJ}$. There is an enhancement for $\omega \chi_{c2}$
around 4.42 GeV, which doesn't appear in the $\omega \chi_{c0,1}$
channels. A coherent sum of the $\psi(4415)$ BW function and a
phase-space term can well describe the $\omega \chi_{c2}$ line shape,
and the branching fraction $\mathcal{B}(\psi(4415)\to\omega\chi_{c2})$
is found to be in the order of $10^{-3}$. The cross section of
$\EE \too \omega \chi_{c1}$ seems to be rising near 4.6 GeV. The
$\omega \chi_{c0}$ is refitted with the higher energy points included,
and the fit results remain almost unchanged. The different line shapes
observed for $\omega\chi_{cJ}$ might indicate that the production
mechanism is different, and that nearby resonances (e.g. $\psi(4415)$)
have different branching fractions to the $\omega\chi_{cJ}(J=0,1,2)$ decay
modes. Further studies based on more data samples at higher energy
will be helpful to clarify the nature of charmonium(-like) states in
this region.

The BESIII collaboration thanks the staff of BEPCII and the IHEP
computing center for their strong support. This work is supported in
part by National Key Basic Research Program of China under Contract
No. 2015CB856700; National Natural Science Foundation of China (NSFC)
under Contracts Nos. 11125525, 11235011, 11322544, 11335008, 11425524;
the Chinese Academy of Sciences (CAS) Large-Scale Scientific Facility
Program; Joint Large-Scale Scientific Facility Funds of the NSFC and
CAS under Contracts Nos. 11179007, U1232201, U1332201; CAS under
Contracts Nos. KJCX2-YW-N29, KJCX2-YW-N45; 100 Talents Program of CAS;
INPAC and Shanghai Key Laboratory for Particle Physics and Cosmology;
German Research Foundation DFG under Contract No. Collaborative
Research Center CRC-1044; Istituto Nazionale di Fisica Nucleare,
Italy; Ministry of Development of Turkey under Contract
No. DPT2006K-120470; Russian Foundation for Basic Research under
Contract No. 14-07-91152; U. S. Department of Energy under Contracts
Nos. DE-FG02-04ER41291, DE-FG02-05ER41374, DE-FG02-94ER40823,
DESC0010118; U.S. National Science Foundation; University of Groningen
(RuG) and the Helmholtzzentrum fuer Schwerionenforschung GmbH (GSI),
Darmstadt; WCU Program of National Research Foundation of Korea under
Contract No. R32-2008-000-10155-0.

\end{document}